\documentclass[journal]{IEEEtran}
\IEEEoverridecommandlockouts

\usepackage{textcomp}
\usepackage[T1]{fontenc}
\usepackage[utf8]{inputenc}
\usepackage{lmodern}
\usepackage{flushend}

\usepackage{amsmath,amssymb,amsfonts,bm}
\usepackage{algorithmic}
\usepackage[ruled,vlined]{algorithm2e}

\usepackage{pgfplots}
\usepackage{grffile}
\pgfplotsset{compat=newest}
\usetikzlibrary{math,calc,plotmarks,arrows.meta}
\usetikzlibrary{arrows.meta}
\usepgfplotslibrary{patchplots}
\usepackage{dblfloatfix}    
\usepackage{graphicx}
\usepackage{caption}
\usepackage{subcaption}
\usepackage{xcolor}
\usepackage{pdfpages}

\usepackage{cite}
\usepackage{hyperref}

\renewcommand*\familydefault{\sfdefault}
\def\BibTeX{{\rm B\kern-.05em{\sc i\kern-.025em b}\kern-.08em
    T\kern-.1667em\lower.7ex\hbox{E}\kern-.125emX}}
\SetKwInput{KwInput}{Input}                
\SetKwInput{KwOutput}{Output}              

\DeclareMathOperator*{\argmax}{arg\,max}
\DeclareMathOperator*{\R}{\mathbb{R}}
\DeclareMathOperator*{\E}{\mathbb{E}}
\renewcommand{\Vec}{\mathbf}

\begin{document}
\title{Unsupervised EEG-based decoding of absolute auditory attention with canonical correlation analysis\\
\thanks{N. Heintz and A. Bertrand are with KU Leuven, Department of Electrical Engineering (ESAT), STADIUS Center for Dynamical Systems, Signal Processing and Data Analytics and also with Leuven.AI - KU Leuven institute for AI, Kasteelpark Arenberg 10, B-3001 Leuven, Belgium (e-mail: nicolas.heintz@esat.kuleuven.be, alexander.bertrand@esat.kuleuven.be). 

N. Heintz  and T. Francart are with KU Leuven, Department of Neurosciences, Research Group ExpORL, Herestraat 49 box 721, B-3000 Leuven, Belgium (e-mail: tom.francart@kuleuven.be).

This research is funded by the Research Foundation - Flanders (FWO) project nr. G0A4918N, the European Research Council (ERC) under the European Union’s Horizon 2020 research and innovation programme (grant agreement No 802895 and grant agreement No 637424), and the Flemish Government (AI Research Program).
The scientific responsibility is assumed by its authors.
}
}

\author{Nicolas~Heintz, Tom~Francart, and~Alexander~Bertrand,~\IEEEmembership{Senior Member, IEEE}}

\maketitle
\begin{abstract}
We propose a fully unsupervised algorithm that detects from encephalography (EEG) recordings when a subject actively listens to sound, versus when the sound is ignored. This problem is known as absolute auditory attention decoding (aAAD). We propose an unsupervised discriminative CCA model for feature extraction and combine it with an unsupervised classifier called minimally informed linear discriminant analysis (MILDA) for aAAD classification. Remarkably, the proposed unsupervised algorithm performs significantly better than a state-of-the-art supervised model. A key reason is that the unsupervised algorithm can successfully adapt to the non-stationary test data at a low computational cost. This opens the door to the analysis of the auditory attention of a subject using EEG signals with a model that automatically tunes itself to the subject without requiring an arduous supervised training session beforehand. 
\end{abstract}


\section{Introduction}
\label{sec:Introduction}
As the sound travels through the ear, it transforms into an electrical stimulus that eventually reaches the brain. The brain then processes this electrical stimulus through a complex cascade of electrical responses, which can be measured using electroencephalography (EEG) \cite{Power2011,Ding2012,Mesgarani2012}.

For amplitude-modulated sounds, such as speech or music, the cortical response tends to track the energy envelope of the incoming sound, which can be quantified using the correlation between the EEG and the speech envelope. This correlation is significantly higher when the subject actively listens to the incoming sound compared to when the subject ignores it \cite{Vanthornhout2019a,roebben_are_2024}. This makes it possible to decode from EEG signals to what degree a person understands and actively listens to a given sound (either or not in the presence of other competing sound sources) \cite{OSullivan2014, Mirkovic2015, Miran2018, Ciccarelli2019, DeCheveigne2018, Monesi2020, Puffay2023, roebben_are_2024}. Such decoders have a plethora of potential applications. For example, in audiology, the decoders yield objective markers of speech intelligibility \cite{Ding2013, Vanthornhout2018, Lesenfants2019}. Moreover, these decoders can also be used to improve contemporary hearing aids with the introduction of adaptive and automatic hearing aid fitting \cite{Vanheusden2020} or can control a noise reduction algorithm to amplify the attended speaker in a multi-talker scenario \cite{OSullivan2014,DeCheveigne2018,Ciccarelli2019,VanEyndhoven2017,Geirnaert2021Unsup}. Finally, Roebben et al. recently demonstrated that such decoders can provide an accurate estimate of the extent to which a person actively pays attention to a given speaker \cite{roebben_are_2024}. 

Many of these decoders are based on regression algorithms that model the interaction between attended sound and measured EEG signals \cite{OSullivan2014, Mirkovic2015, Miran2018, Ciccarelli2019, Taillez2020, DeCheveigne2018, Monesi2020}. This interaction is different for each subject \cite{Mirkovic2015} and for different listening scenarios \cite{Fuglsang2017}. Although it is possible to train models that can generalise over multiple subjects and conditions, this leads to a substantial decrease in performance \cite{Mirkovic2015, OSullivan2014, Geirnaert2020}. Alternatively, the model can be calibrated for each subject separately during a calibration session, although such a time-consuming practice would be impractical in real-life applications. Moreover, such subject-specific decoders are fixed and inherently assume that the neural response is stationary in time. This is not true in reality. Over time, the cortical response can change, electrodes can move, deteriorate and fail, the conductance between the electrodes and skin can vary, etc. 

More and more brain-computer interfaces, therefore, make use of adaptive models \cite{Lotte2018a, Geirnaert2022Unsup, Beauchene2023}. These models are continuously retrained on the incoming data, which allows them to adapt to the changes mentioned earlier. Since they can even be initialised with a random model, there is no need for a lengthy calibration session \cite{Geirnaert2022Unsup}. However, hearing devices are generally used outside a supervised lab environment, where there are no ground-truth labels indicating whether/to whom the subject is listening. In contrast, most state-of-the-art algorithms decoding auditory attention require such ground-truth labels to be trained \cite{Geirnaert2020}. 

This problem was first tackled in \cite{Geirnaert2021Unsup, Geirnaert2022Unsup}, where an unsupervised linear regression model decodes to which of multiple competing speaker a subject is listening. This method was improved in \cite{Heintz2023Unbiased}, which corrected a bias in the original algorithm that hampered convergence on smaller datasets. However, \cite{Geirnaert2021Unsup, Geirnaert2022Unsup, Heintz2023Unbiased} all exclusively focused on selective Auditory Attention Decoding (sAAD, often simply called AAD), where the algorithm decodes to which of multiple competing speakers a subject is listening. 

This paper proposes an unsupervised algorithm that tackles the absolute Auditory Attention Decoding (aAAD) problem. Here, the objective is to decode whether or not a subject is actively paying attention to a single speaker \cite{roebben_are_2024, Vanthornhout2019a, lesenfants_interplay_2020}. The proposed model is based on the state-of-the-art \textit{supervised} canonical correlation analysis (CCA) algorithm proposed in \cite{DeCheveigne2018, Cheveigne2020} for feature extraction, though adapted to make it more discriminative and unsupervised. These features are then classified using an unsupervised variant of linear discriminant analysis called minimally informed linear discriminant analysis (MILDA) \cite{heintz_minimally_2023} and thresholded using a Gaussian mixture model (GMM). 

Remarkably, the proposed unsupervised algorithm significantly outperforms the state-of-the-art supervised CCA model, even when the unsupervised algorithm is initialised randomly. This has two complementary causes: 
\begin{itemize}
    \item EEG is highly non-stationary. The unsupervised algorithm is automatically tuned to the changing statistics of the incoming unlabelled test data, while this is not possible for supervised algorithms. 
    \item The ground-truth labels are imperfect. It is easy for a subject to briefly pay attention to the incoming sound when they shouldn't and vice versa. This negatively affects the performance of the supervised model, which uses these imperfect labels \cite{roebben_are_2024}.
\end{itemize}

The focus on a CCA+LDA model is motivated by its strong performance for match-mismatch and AAD tasks compared to other linear models \cite{Cheveigne2020}. Furthermore, the linearity allows for an efficient updating strategy with low computational cost, which makes them perfect for use in a wearable setting, where adaptive models are the most relevant.

The paper is structured as follows. We first explain the supervised CCA algorithm on which our model is based in Section \ref{sec:Supervised}. We then convert this model into an unsupervised self-adaptive model in Section \ref{sec:Unsupervised}. Section \ref{sec:Experiments} discusses the experimental procedures and data that are used to validate our algorithm. The results of these experiments are discussed in Section \ref{sec:Results}. Finally, suggestions for future work and the conclusion can be found in Section \ref{sec:Conclusion}.

\section{Supervised absolute Auditory Attention Decoding} \label{sec:Supervised}
In an absolute Auditory Attention Decoding (aAAD) task, we wish to classify whether a subject actively listens to a specific speaker at each moment in time \cite{roebben_are_2024}. In this Section, we explain how such a problem can be tackled in a supervised way using discriminative CCA (Section \ref{subsec:SupCCA}) and Fisher's LDA (Section \ref{subsec:SupLDA}). The unsupervised algorithm is explained in Section \ref{sec:Unsupervised}. 

\subsection{CCA-based feature extraction} \label{subsec:SupCCA}
Consider a C-channel EEG signal of which $\Vec{x}(t) \in \R^{C \times 1}$ denotes the EEG signal at sample time $t$. The c\textsuperscript{th} channel of $\Vec{x}(t)$ is denoted as $x_c(t)$. Assume the EEG is recorded while a subject can hear some speech signal with an envelope\footnote{The envelope $s(t)$ is assumed to capture the short-term energy or amplitude modulations (typically within the 0-10Hz range) of the speech and can be defined and computed in various ways \cite{Biesmans2017} (see also Section \ref{sec:Experiments}).} $s(t)$. The subject only actively listens to the speech signals when $t \in \mathcal{T_+}$ and ignores the speech when $t \in \mathcal{T_-}$. 

Classically, the CCA module looks for a joint (linear) spatio-temporal transformation of the EEG and speech segments, such that they are maximally correlated to each other in the transformation space  when the subject is actively listening. 

Formally, this transformation can be expressed as:
\begin{equation}
\begin{split}
    & \Bar{x}(t) = \sum_{\tau = 0}^{L_x-1} \sum_{c=1}^C x_c(t-\tau+S) D(\tau,c) \\
    & \Bar{s}(t) = \sum_{\tau = 0}^{L_s-1}s(t-\tau) e(\tau).
\end{split}
\label{eq:XmatrixDecod}
\end{equation}
In this equation, the speech envelope is transformed with a finite impulse response filter $\Vec{e} \in \R^{L_s \times 1}$. The EEG signal is transformed with a spatio-temporal decoder $D \in \R^{L_x \times C}$. The EEG signal and envelope are lagged with $L_x$ and $L_s$ samples respectively to construct the temporal filters. The EEG signal is also delayed with an additional $S$ samples relative to the speech envelope to correct for the fact that a cortical response can only occur after hearing the speech and thus always lags behind the speech envelope. 

To ease notation, the $L_x$ time lags of all $C$ EEG channels are typically combined in a time-dependent vector $\Vec{x}(t) \in \R^{CL_x}$:
\begin{equation}
\begin{split}
    &\Vec{x}(t) = [\Vec{x}_1(t)^T \hdots \Vec{x}_C(t)^T]^T\\
    &\Vec{x}_c(t) = [x_c(t-L_x+S+1) \hdots x_c(t+S)]^T.
\end{split}
\label{eq:xVector}
\end{equation}

Similarly, the lagged speech envelope can be written as
\begin{equation}
    \Vec{s}(t) = [s(t-L_s+1) \hdots s(t)]^T,
    \label{eq:sVector}
\end{equation}
which allows us to simplify the equations in \eqref{eq:XmatrixDecod} to: 

\begin{equation}
\begin{split}
    &\Bar{x}(t) = \Vec{d}^T\Vec{x}(t)\\
    &\Bar{s}(t) = \Vec{e}^T\Vec{s}(t). 
\end{split}
\end{equation}

The vectors $\Vec{d}$ and $\Vec{e}$ are typically chosen such that the Pearson correlation coefficient between $\Bar{x}(t)$ and $\Bar{s}(t)$ is maximised when the subject is actively listening ($t\in \mathcal{T_+}$) \cite{DeCheveigne2018,Cheveigne2020}:
\begin{equation}
\begin{split}
\Vec{d},\Vec{e}=& \argmax_{\Vec{d},\ \Vec{e}} \frac{\E[\Vec{d}^T\Vec{x}(t)\Vec{s}(t)^T\Vec{e},\ t\in\mathcal{T_+}]}{\sqrt{\E[\Vec{d}^T\Vec{x}(t)\Vec{x}(t)^T\Vec{d}]}\sqrt{\E[\Vec{e}^T\Vec{s}(t)\Vec{s}(t)^T\Vec{e}]}}\\
=& \argmax_{\Vec{d},\ \Vec{e}} \frac{\Vec{d}^TR_{xs,+}\Vec{e}}{\sqrt{\Vec{d}^TR_{xx}\Vec{d}}\sqrt{\Vec{e}^TR_{ss}\Vec{e}}},
\end{split}
\label{eq:CCAOptimProblem}
\end{equation}
where $\E[.]$ denotes the expectation operator, $R_{xs,+} = \E[\Vec{x}(t)\Vec{s}(t)^T,\ t\in\mathcal{T_+}] \in \R^{CL_x\times L_s}$ is the cross-correlation matrix between the EEG signal and the speech envelope when a subject is actively listening, and where $R_{xx} = \E[\Vec{x}(t)\Vec{x}(t)^T] \in \R^{CL_x\times CL_x}$ and $R_{ss}= \E[\Vec{s}(t)\Vec{s}(t)^T]\in \R^{L_s \times L_s}$ are the autocorrelation matrices of the EEG signal and the speech envelope respectively.

However, in aAAD the goal is to maximally discriminate between attended and unattended speech, rather than exclusively maximising attended speech. We, therefore, also propose an alternative called discriminative CCA (dCCA), where the difference in correlation between the attended and unattended classes is maximised:
\begin{equation}
\begin{split}
\Vec{d},\Vec{e}=& \argmax_{\Vec{d},\ \Vec{e}} \frac{\Vec{d}^TR_{xs,+}\Vec{e}-\Vec{d}^TR_{xs,-}\Vec{e}}{\sqrt{\Vec{d}^TR_{xx}\Vec{d}}\sqrt{\Vec{e}^TR_{ss}\Vec{e}}}\\
&= \argmax_{\Vec{d},\ \Vec{e}} \frac{\Vec{d}^TR_{xs,\Delta}\Vec{e}}{\sqrt{\Vec{d}^TR_{xx}\Vec{d}}\sqrt{\Vec{e}^TR_{ss}\Vec{e}}},
\end{split}
\label{eq:dCCAOptimProblem}
\end{equation}
with $R_{xs,\Delta} = R_{xs,+}-R_{xs,-}$ the difference in the cross-correlation matrix between EEG and speech for attended and unattended speech.

The solutions of this optimisation problem are the eigenvectors with the largest eigenvalue $\lambda$ of the following generalised eigenvalue problems:
\begin{equation}
\begin{cases}
    &R_{xs}R_{ss}^{-1}R_{xs,\Delta}^T\Vec{d} = \lambda R_{xx}\Vec{d}\\
    &R_{xs}^TR_{xx}^{-1}R_{xs,\Delta}\Vec{e} = \lambda R_{ss}\Vec{e}.
\end{cases}
\label{eq:jointEigCCA}
\end{equation}

It is nevertheless possible to compute more than just one unique transformation. The equations in \eqref{eq:jointEigCCA} have at most $L_s$ linearly independent solutions, assuming that $R_{xx}$ and $R_{ss}$ are of full rank and $CL_x \geq L_s$. The eigenvectors $\Vec{d}_2,\ \Vec{e}_2$ with the second largest $\lambda$ give a transformation with maximal correlation orthogonal to the first transformation, etc. The CCA model can thus compute $K\leq L_s$ different transformations, which together capture orthogonal signal components that are maximally correlated between the EEG and the speech envelope \cite{DeCheveigne2018}. By correlating the transformed EEG and speech over a finite segment length after each of the $K$ transformations, the CCA model thus composes a feature vector $\bm{\rho} \in \R^{K \times 1}$ containing $K$ correlation features. This feature vector is then classified by the LDA classifier explained in the following section. The length of the segment over which the correlations in $\bm{\rho}$ are computed corresponds to the amount of data the algorithm uses to decide whether or not a subject is actively listening. Therefore, we refer to the segment length as the 'decision window length' in the remainder of this paper. Longer decision windows lead to less noisy estimates of $\bm{\rho}$, and therefore generally to higher decoding accuracies at the cost of a worse temporal accuracy.

\subsection{Classification} \label{subsec:SupLDA}
The feature vector $\bm{\rho}$ computed in Section \ref{subsec:SupCCA} is classified with Fisher's LDA using a classification vector $\Vec{w} \in \R^{K \times 1}$ that maximises the scatter between classes and minimises the scatter within classes \cite{Bishop2006}:

\begin{equation}
    \tilde{\Vec{w}} = \argmax_\Vec{w} \frac{\left(\Vec{w}^T(\bm{\mu}_+-\bm{\mu}_-)\right)^2}{\Vec{w}^T(\Sigma_+ + \Sigma_-)\Vec{w}}.
    \label{eq:FLDAObj}
\end{equation}
It is easy to show that 
\begin{equation}
    \Vec{w} \propto (\Sigma_+ + \Sigma_-)^{-1}(\bm{\mu}_+-\bm{\mu}_-)
    \label{eq:FLDA}
\end{equation}
maximises \eqref{eq:FLDAObj} \cite{Bishop2006}. $\bm{\rho}_n$ is predicted to belong to the match class if $y_n = \Vec{w}^T\bm{\rho}_n>T$, with $T$ some threshold.


\section{Unsupervised absolute Auditory Attention Decoding} \label{sec:Unsupervised}
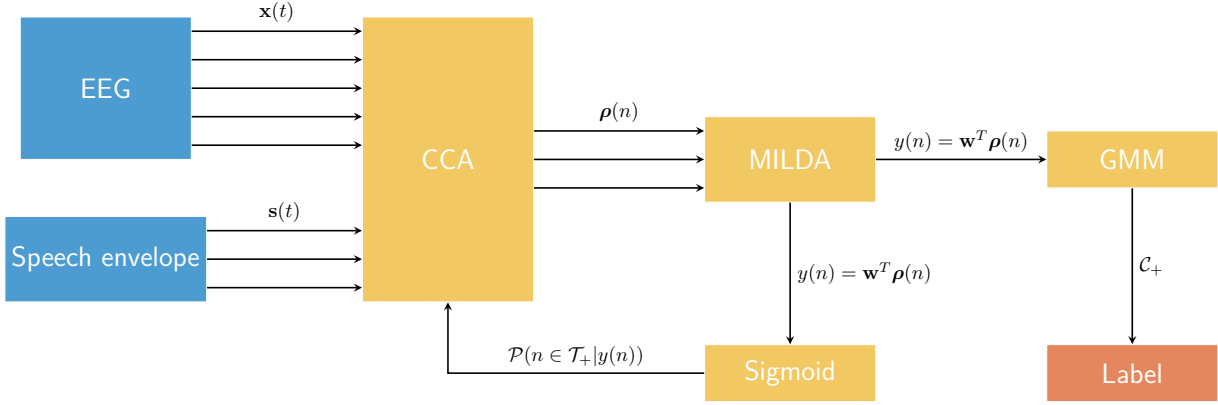
\begin{figure*}[t]
    \centering
    \usetikzlibrary{shapes,arrows,fit,calc,positioning,automata}

\definecolor{mycolor1}{rgb}{0.00000,0.44700,0.74100}%
\definecolor{mycolor2}{rgb}{0.85000,0.32500,0.09800}%
\definecolor{mycolor3}{rgb}{0.92900,0.69400,0.12500}%

\tikzstyle{input} = [rectangle, minimum width=3cm,text centered,  text = white,draw=white, fill=mycolor1!70]
\tikzstyle{io} = [rectangle, minimum width=3cm, minimum height=1cm,text centered,  text = white,draw=white, fill=mycolor3!70]
\tikzstyle{cca} = [minimum width=3cm, minimum height=5cm,text centered, text = white, draw=white, fill=mycolor3!70]
\tikzstyle{output} = [rectangle, minimum width=3cm,minimum height = 1cm, text centered,  text = white, draw=white, fill=mycolor2!70]
\tikzstyle{arrow} = [thick,->,>=stealth]

\resizebox{0.9\textwidth}{!}{
\begin{tikzpicture}[node distance = 6cm]
\node (EEG) [input, minimum height = 2.5cm] {\Large EEG};
\node (Stim) [input, minimum height = 1.5cm, below of=EEG, yshift = 3cm] {\Large Speech envelope};
\node (CCA) [cca, right of=EEG, yshift = -1.25cm] {\Large CCA};
\node (uLDA) [io, right of=CCA, minimum height = 1.5cm] {\Large MILDA};
\node (sigmoid) [io, below of=uLDA, yshift = 2.25cm] {\Large Sigmoid};
\node (thres) [io, right of=uLDA] {\Large GMM};
\node (lab) [output, below of=thres, yshift = 2.25cm] {\Large Label};

\foreach \i [count=\xi from 0] in {2,...,-2}{%
  \draw[arrow] ([yshift=\i * 0.5 cm]EEG.east) -- ([yshift=1.25 cm + \i * 0.5 cm]CCA.west) node[midway] (o\xi) {} ;}
 
\node[above = -0.1cm of o0] (d3) {$\Vec{x}(t)$};

\foreach \i [count=\xi from 0] in {1,...,-1}{%
  \draw[arrow] ([yshift=\i * 0.5 cm]Stim.east) -- ([yshift=-1.75 cm + \i * 0.5 cm]CCA.west) node[midway] (o\xi) {} ;}
 
\node[above = -0.1cm of o0] (d3) {$\Vec{s}(t)$};

\foreach \i [count=\xi from 0] in {1,...,-1}{%
  \draw[arrow] ([yshift = \i * 0.5 cm]CCA.east) -- ([yshift=\i * 0.5 cm]uLDA.west) node[midway](o\xi){};}
 
\node[above = -0.1cm of o0] (d3) {$\bm{\rho}(n)$};

\draw[arrow] (uLDA.east) -- (thres.west) node[above,midway]{$y(n)=\Vec{w}^T\bm{\rho}(n)$};
\draw[arrow] (thres.south) -- (lab.north) node[right,midway]{$\mathcal{C}_+$};
\draw[arrow] (uLDA.south) -- (sigmoid.north) node[right,midway]{$y(n)=\Vec{w}^T\bm{\rho}(n)$};
\draw[arrow] (sigmoid.west) -| (CCA.south) node[pos=0.25,above]{$\mathcal{P}(n\in \mathcal{T_+}|y(n))$};
\end{tikzpicture}
}
    \caption{A summary of the unsupervised aAAD algorithm. Blue boxes reflect inputs, yellow boxes contain the modules of the actual algorithm and red boxes the output.}
    \label{fig:unsupSummary}
\end{figure*}

In an unsupervised aAAD model, there is no knowledge when a subject is actively paying attention to the sound. This information is needed for both the training of the CCA feature extraction and the LDA classifier. Due to the absence of such labels, we propose to iteratively optimise the CCA model using an Expectation-Maximisation approach. At each iteration, the CCA model is re-trained using soft labels that estimate the probability $\mathcal{P}(t\in \mathcal{T_+})$ that the subject was actively listening to the speech one time $t$ using the output of the CCA model and classifier from the previous iteration.

Such an iterative process relies on a self-leveraging effect: better predictions should lead to better models and, thus, even better predictions. However, such a self-leveraging effect is not always present and can also have an adverse effect on the performance. Consider a case where the initial predictions are completely correct. Since the trained (in this case supervised) aAAD model is not perfect, it will make less correct predictions in the next iteration. These less accurate predictions are then used to train a worse model in the second iteration, which thus makes even worse predictions. This process continues until it converges to some equilibrium point where models trained in consecutive iterations have exactly the same performance. Note that the performance in this steady-state regime is not necessarily better than chance.

In this section, we will explore the self-leveraging effect in an aAAD model and adapt the algorithm explained in Section \ref{sec:Supervised} such that the lack of ground-truth labels has a minimal effect on the performance. We will argue that the self-leveraging effect is naturally present in the ordinary CCA model, but not necessarily in the discriminative CCA model or classifier. Therefore, we will modify the classification step to facilitate an unsupervised match-mismatch classification. An overview of the unsupervised algorithm is shown in Figure \ref{fig:unsupSummary} and in Algorithm \ref{alg:UnsupCCA}.

\subsection{CCA-based feature extraction} \label{subsec:UnsupFeatExtr}
First, we consider the ordinary formulation of CCA proposed in \eqref{eq:CCAOptimProblem} where only the correlation between EEG and attended speech is maximised. In order to compute an optimal decoder $\Vec{d}$ and encoder $\Vec{e}$, three correlation matrices must be correctly estimated: $R_{xx}$, $R_{ss}$ and $R_{xs,+}$.

\begin{itemize}
    \item $R_{xx}$ is the autocorrelation of the EEG signal. While it can differ when a subject is or isn't paying attention, such spatial features are known to be unreliable \cite{roebben_are_2024}. Therefore, the proposed models estimate $R_{xx}$ on all recorded EEG model and does not require any labels. 
    \item $R_{ss}$ is the autocorrelation of the speech envelope. As speech statistics are universal, they do not depend on aAAD labels. This matrix can thus be estimated with all available speech segments. 
    \item $R_{xs,+}$ is the cross-correlation between EEG and attended speech, i.e., it captures the correlation between the speech stimulus and the stimulus-following response in the EEG. Since unattended speech should not be included in the estimation, $R_{xs,+}$ requires labels for a good estimation. 
\end{itemize}

Only the estimation of the cross-correlation matrix $R_{xs,+}$ depends on the predictions from the previous iteration. If these predictions are imperfect, the estimated $\Bar{R}_{xs,+}$ is a mixture of $R_{xs,+}$ and $R_{xs,-}$. If the previous predictions were random and the dataset is balanced, $\Bar{R}_{xs,+}\propto R_{xs,+}+ R_{xs,-}$. In this case the CCA model tries to maximise the correlation of the attended and unattended class equally, which obviously decreases the separation between these two classes. The objection function is in this case: 

\begin{equation}
\Vec{d},\Vec{e}= \argmax_{\Vec{d},\ \Vec{e}} \frac{\Vec{d}^TR_{xs,+}\Vec{e}+\Vec{d}^TR_{xs,-}\Vec{e}}{\sqrt{\Vec{d}^TR_{xx}\Vec{d}}\sqrt{\Vec{e}^TR_{ss}\Vec{e}}}.
\label{eq:chanceLevelObjective}
\end{equation}
However, even if both correlations are equally maximised in the objective function, there will still be a difference between the attended and unattended class. This is because EEG is naturally more correlated to attended speech than unattended speech \cite{Vanthornhout2019a}. Therefore, \eqref{eq:chanceLevelObjective} will put more weight on $\Vec{d}^TR_{xs,+}\Vec{e}$ than on $\Vec{d}^TR_{xs,-}\Vec{e}$, causing the attended correlations to be (slightly) larger than the unattended correlations on average. This causes the predictions in the next iteration to be better than chance, hence causing a self-leveraging effect. This effect is further explored in \cite{Geirnaert2021Unsup} and experimentally validated in Section \ref{sec:Results}.

This self-leveraging effect is not present for the discriminative CCA proposed in \eqref{eq:dCCAOptimProblem}. This is because a wrong prediction directly counteracts the initial objective. Indeed, if all predictions are random, half the terms will try to maximise the difference between the attended and unattended correlations, and the other half will try to minimise it (i.e., maximise the negative). This causes the iterative procedure to converge quickly to a suboptimal point. Since discriminative CCA outperforms normal CCA when high-quality predictions are available, as shown in Section \ref{sec:Experiments}, we propose to use normal CCA in Algorithm \ref{alg:UnsupCCA} for the iterative procedure until it obtains high-quality predictions and exclusively use discriminative CCA in the last iteration. 

\subsection{Classification} \label{subsec:unsuperClass}
The CCA model produces a $K$-dimensional feature vector $\bm{\rho}(n)$ containing the correlations between the projected EEG $\Vec{x}(t)$ and speech envelope $\Vec{s}(t)$ during a time window $n$ in a latent space produced by $K$ orthogonal pairs of decoders $d$ and encoders $e$. The goal of the classification module is to project these $K$ correlations onto a 1-dimensional score that separates attended and unattended correlations as well as possible. 

Similar to discriminative CCA, LDA does not have a reliable self-leveraging effect. Moreover, classical unsupervised clustering algorithms perform poorly in aAAD due to the large class variability. Instead, an unsupervised classifier called minimally informed linear discriminant analysis (MILDA) is used. This unsupervised model is equivalent to LDA if the class averages of the features are proportional to each other, i.e., $\bm{\mu_+} \propto \bm{\mu_-}$, with $\bm{\mu_\pm} \triangleq \frac{1}{|\mathcal{T_\pm}|} \sum_{t\in\mathcal{T_\pm}}\rho(n)$ the class averages of the attended and unattended class \cite{heintz_minimally_2023}. If this condition is satisfied, and $\Bar{\bm{\mu}}\neq \Vec{0}$, the MILDA projection vector

\begin{equation}
    \begin{split}
    \Vec{w} &= \Bar{\Sigma}^{-1}\Bar{\bm{\mu}}\\
        &\Bar{\bm{\mu}} = \frac{1}{N} \sum_{t} \bm{\rho}(n)\\
        &\Bar{\Sigma}= \frac{1}{N} \sum_{t} (\bm{\rho}(n)-\Bar{\bm{\mu}})(\bm{\rho}(n)-\Bar{\bm{\mu}})^T,
    \end{split}
\end{equation} 
is equivalent to the LDA projection in \eqref{eq:FLDA} \cite{heintz_minimally_2023}. 

Though it is not necessarily true that $\bm{\mu_+} \propto \bm{\mu_-}$ in this use case, it is possible to transform the features such that this condition is satisfied using some mild assumptions. Below, we will show that the direction of largest variance is expected to be identical to the difference between the two class averages $\bm{\mu_\Delta}$. Once $\bm{\mu_\Delta}$ is known, it is relatively straightforward to transform the feature space such that $\bm{\mu_+} \propto \bm{\mu_-}$ and $\Bar{\bm{\mu}}\neq \Vec{0}$. 

Assume that the EEG $\Vec{x}(t)$ recorded when a subject pays attention to sound with envelope $\Vec{s}(t)$ consists of three components: the neural response caused by paying attention to the sound $\alpha(t) A\Vec{s}(t)$, the neural response caused by hearing the sound $B\Vec{s}(t)$ and a (typically large) noise component $\Vec{n}(t)$ that is linearly uncorrelated with the speech envelope. $\alpha(t)$ scales the attention term depending on how attentive the subject is. Further, assume w.l.o.g. that the EEG and speech envelopes were scaled such that $\sum_{t\in\mathcal{T}_n}\Vec{d}^T\Vec{x}(t)\Vec{x}(t)^T\Vec{d}=\sum_{t\in\mathcal{T}_n}\Vec{e}^T\Vec{s}(t)\Vec{s}(t)^T\Vec{e}=1$ in each window $n$.

The correlation $\rho_k(n)$ between the projected EEG and speech envelope with decoder $\Vec{d}_k$ and encoder $\Vec{e}_k$ in a specific window $n$ where $t\in\mathcal{T}_n$ is: 
\begin{equation}
    \begin{split}
        &\rho_k(n) \\
        &= \frac{1}{|\mathcal{T}_n|}\sum_{t\in\mathcal{T}_n}\Vec{d}_k^T\Vec{x}(t)\Vec{s}(t)^T\Vec{e}_k\\
         &= \frac{1}{|\mathcal{T}_n|}\sum_{t\in\mathcal{T}_n}\Vec{d}_k^T(\alpha(t) A+B)\Vec{s}(t)\Vec{s}(t)^T\Vec{e}_k +\Vec{d}_k^T\Vec{n}(t)\Vec{s}(t)^T\Vec{e}_k\\
         &\approx \Vec{d}_k^T(\alpha(n) A+B)R_{ss}(n)\Vec{e}_k +\Vec{d}_k^TR_{ns}(n)\Vec{e}_k\\
         &= \Vec{d}_k^T(\alpha(n) A+B)R_{ss}\Vec{e}_k \\
         &+\Vec{d}_k^T(\alpha(n) A+B)(R_{ss}(n)-R_{ss})\Vec{e}_k\\
         &+\Vec{d}_k^TR_{ns}(n)\Vec{e}_k,
    \end{split}
    \label{eq:corrFeatureModel}
\end{equation}
with $R_{ss}(n)$ the sample estimation of $R_{ss}$ using the available data in window $n$.

Since $\E[R_{ns}(n)]=0$, the class averages of $\rho_k(n)$ are:
\begin{equation}
    \begin{split}
        \mu_{k,+} &= \Vec{d}_k^T(\Bar{\alpha}_+A+B)R_{ss}\Vec{e}_k\\
        \mu_{k,-} &= \Vec{d}_k^T(\Bar{\alpha}_-A+B)R_{ss} \Vec{e}_k,
    \end{split}
\end{equation}
with $\Bar{\alpha}_+, \Bar{\alpha}_-$ the average degree of attention paid in the attended and unattended classes, respectively. The difference between the two class averages is $\bm{\mu}_\Delta$, with:

\begin{equation}
     \mu_{\Delta,k}= \Vec{d}_k^T(\Bar{\alpha}_+-\Bar{\alpha}_-)R_{ss}\Vec{e}_k.
\end{equation}

As shown in \eqref{eq:corrFeatureModel}, the correlation features have three important sources of variance: 
\begin{itemize}
    \item $\alpha(n)$: the variance caused by changes in attention by the subject. This is the source of variance that causes a difference between the attended and unattended class and has the direction $\bm{\mu}_\Delta$. This variance is maximised in discriminative CCA, but also observed to be large in normal CCA. 
    \item $R_{ss}(n)-R_{ss}$: deviations on the autocorrelation of the envelope. The deviation of $R_{ss}(n)$ from $R_{ss}$ is quite random. As a consequence, $\Vec{d}_k^T(\alpha(n) A+B)(R_{ss}(n)-R_{ss})\Vec{e}_k$ lacks any clear structure is a completely unrelated to $\Vec{d}_k$ and $\Vec{e}_k$. Since $\|\Vec{d}_k\|=\|\Vec{e}_k\|=1\  \forall k$, this variation is thus expected to have a similar impact on all $k$ correlation features.
    \item $R_{ns}(n)$: the correlation between the random noise and the speech envelope. Similar to $R_{ss}(n)-R_{ss}$, this matrix is random and on average 0. The variance caused by the correlation features is thus also expected to be similar for all $k$ correlation features.
\end{itemize}
The variation in attention is thus the only source of variance that is directed in the feature space. Since all other sources of variance are expected to be equally large in all directions, the largest direction of variance in the complete feature space is thus expected to be proportional to $\bm{\mu}_\Delta$. Since the direction of largest variance is simply the eigenvector corresponding to the largest eigenvalue of $\Bar{\Sigma}$, it is possible to find the vector $\bm{\delta} \propto \bm{\mu}_\Delta$ without labels. 

Once $\bm{\delta}$ is known, the transformation $\bm{\rho}'(n) = \bm{\rho}(n) - \Bar{\bm{\mu}} + \delta$ ensures that the class averages of the transformed features $\bm{\rho}'(n)$ are proportional to each other, satisfying the sole requirement for MILDA to be equivalent to LDA \cite{heintz_minimally_2023}.

\subsection{Thresholding} \label{subsec:unsupThres}
The classification module projected the $K-$dimensional feature vectors $\bm{\rho}(n)$ onto 1-dimensional scores $y(n)$. These scores suffice to create the rough estimations of the probability that the subject is paying attention that are used in the iterative process using the sigmoid transformation $p(n) = 1/(1+e^{-(y(n)-\Bar{y})/\sigma_y}$, with $\Bar{y}$ the average of all scores and $\sigma_y$ their standard deviation. 

However, these probabilities should be estimated with greater care in the last iteration, as these are the final estimated probabilities reported back to the user.  
Automatically determining the optimal threshold is not trivial. A straightforward iterative estimation of the threshold $T$ is suboptimal since this process also lacks the self-leveraging effect. An overestimation of a threshold would lead to an overestimation of $\mu_+$ and $\mu_-$, and thus once more to an overestimation of the threshold in the next iteration. 

We therefore propose the following strategy for the case where $q$ is not known. We model the probability distribution of the one dimensional scores $y_n = \Vec{w}^T\bm{\rho}_n$ with a Gaussian Mixture Model (GMM) containing two normal distributions. The distribution $\mathcal{N}_+(\mu_+,\sigma_+^2)$ corresponds to the match class and the distribution $\mathcal{N}_-(0,\sigma_-^2)$ to the mismatch class. The GMM is fitted on the obtained scores $y_n$ with an Expectation-Maximization algorithm \cite{Bishop2006}, and then used to label each pair of segments as a match or mismatch. A pair of segments $n$ is labelled as a match if the likelihood that $y_n$ is sampled from the match class is larger than the likelihood to be sampled from the mismatch class, i.e. if:

\begin{equation}
    \frac{1}{\sigma_+\sqrt{2\pi}}e^{\frac{-1}{2}\left(\frac{y_n-\mu_+}{\sigma_+}\right)^2}>\frac{1}{\sigma_-\sqrt{2\pi}}e^{\frac{-1}{2}\left(\frac{y_n}{\sigma_-}\right)^2},
\end{equation}
and vice versa. When $\sigma_+ \approx \sigma_-$ (which is expected), this corresponds to a threshold $T=\frac{\mu_+}{2}\approx\frac{\mu_+ + \mu_-}{2}$, which is identical to the threshold used when $q$ is known.

Algorithm \ref{alg:UnsupCCA} and Figure \ref{fig:unsupSummary} show the entire algorithm to label a batch of $N$ pairs of EEG and speech segments with an unknown fraction of matched segments $q$. Note that the algorithm can easily be modified for online applications with a continuous stream of new segments at the input. In those cases, a new segment is first classified by the model and then used to update the unsupervised model according to the updating rules written in Algorithm \ref{alg:UnsupCCA}. 

\begin{algorithm*}

\KwData{$N$ segments of EEG and speech. The segments are of length $\tau$ (corresponding to the decision window length) and are represented as $X_{n} \in \R^{CL_x \times \tau}$ and $S_{n} \in \R^{L_s \times \tau}$, for $n=1...N$, where each column of $X_n$ and $S_n$ is defined as $\Vec{x}(t)$ in \eqref{eq:xVector} and $\Vec{s}(t)$ in \eqref{eq:sVector}, respectively, and where $C$ is the number of EEG channels (or number of PCA components in case of a pre-processing with PCA).}
\KwInput{An initial set of soft labels $p^{(0)}(n) \in [0,1]$, the maximal number of iterations $i_{max}$ and the number of extracted features $K$.}
\KwOutput{A trained model and a predicted label for each segment.}
Compute: 
 \begin{align*}
        &R_{xx} = \frac{1}{N} \sum_{n=1}^N X_nX_n^T\\
        &R_{ss} = \frac{1}{N} \sum_{n=1}^N S_nS_n^T
    \end{align*}
\While{$i \leq i_{max}$}{
    \eIf{$i=i_{max}$}{
    Compute the maximally discriminative CCA model with soft labels
    \begin{equation*}
        R_{xs} = \frac{1}{\sum_n p^{(i)}(n)} \sum_{n=1}^N p^{(i)}(n)X_nS_n^T - (1-p^{(i)}(n)) X_nS_n^T.
    \end{equation*}
    }
    {Compute the normal CCA model with soft labels
    \begin{equation*}
        R_{xs} = \frac{1}{\sum_n p^{(i)}(n)} \sum_{n=1}^N p^{(i)}(n)X_nS_n^T.
    \end{equation*}
    }
   
    Compute the $K$ linearly independent decoders $\Vec{d}_{k}$ and encoders $\Vec{e}_{k}$ by solving the following generalised eigenvalue problem for the $K$ largest eigenvalues:
    
    \begin{equation*}
    \begin{cases}
        &R_{xs}R_{ss}^{-1}R_{xs}^T\Vec{d}_{k} = \lambda R_{xx}\Vec{d}_{k}\\
        &R_{xs}^TR_{xx}^{-1}R_{xs}\Vec{e}_{k} = \lambda R_{ss}\Vec{e}_{k}.
    \end{cases}
    \end{equation*}
    
    Create the feature vectors $\bm{\rho} (n)= [\rho_1(n) \hdots \rho_K(n)]^T \in \R^{K \times 1}$:
    \begin{equation*}
        \rho_k(n) = \frac{\Vec{d}_{k}^TX_nS_n^T\Vec{e}_{k}}{\sqrt{\Vec{d}_{k}^TX_nX_n^T\Vec{d}_{k}}\sqrt{\Vec{e}_{k}^TS_nS_n^T\Vec{e}_{k}}}.
    \end{equation*}
    
    Compute global covariance and average of all feature vectors $\bm{\rho}(n)$, irrespective of their label.
    \begin{equation*}
    \begin{split}
        &\Bar{\bm{\mu}} = \frac{1}{N} \sum_{n=1}^N \bm{\rho}_n\\
        &\Bar{\Sigma}= \frac{1}{N} \sum_{n=1}^N (\bm{\rho}_n-\Bar{\bm{\mu}})(\bm{\rho}_n-\Bar{\bm{\mu}})^T.
    \end{split}
    \end{equation*} 
    
    Compute $\delta$, which is proportional to the eigenvector corresponding to the largest eigenvalue of $\Bar{\Sigma}$.
    
    Compute the scores $y(n)= \delta^T\Bar{\Sigma}^{-1}\bm{\rho}(n)$\\
    Compute $p^{(i+1)}(n) = 1/(1+e^{-(y(n)-\Bar{y})/\sigma_y})$.
    }

    Fit the unattended distribution $\mathcal{N}(\mu_-,\sigma^2)$ and the attended distribution $\mathcal{N}(\mu_+,\sigma^2)$ on the obtained scores $y(n)$ with an Expectation-Maximization algorithm.\\
    Label the segments $\mathcal{C}_+$:
    \begin{equation*}
        \mathcal{C}_+ = \left\{n \in [1,N] \middle| \frac{1}{\sigma\sqrt{2\pi}}e^{\frac{-1}{2}\left(\frac{y(n)-\mu_+}{\sigma}\right)^2} > \frac{1}{\sigma\sqrt{2\pi}}e^{\frac{-1}{2}\left(\frac{y(n)-\mu_-}{\sigma}\right)^2}\right\}.
    \end{equation*}

\caption{Unsupervised match-mismatch classification}
\label{alg:UnsupCCA}
\end{algorithm*}

\section{Experiments}
\label{sec:Experiments}

\subsection{Dataset}\label{subsec:Dataset}
In our experiments, we use the dataset that was first presented in \cite{roebben_are_2024}. The dataset consists of 10 Flemish speaking subjects and contains 4 trials per subject. In each trial, the subject can hear an audiobook narrating different (unknown) Flemish stories at a constant volume. In the first trial, each subject was asked to listen attentively to the story for the entire duration (18 minutes). In the second and third trial, the subject was also instructed to attentively listen to the story, apart from several blocks of 1 minute where they had to ignore the story and solve as many mathematical exercises as possible instead. The second trial contains 25 minutes of data in total, in which the subject was distracted with mathematics for 3 blocks of 1 minute. The third trial consists of 15 minutes, with 2 1-minute blocks of mathematics. In the fourth trial, each subject was instructed to ignore the audiobook and read a different story for the entire duration of 13 minutes. In total, the datasets thus contains 53 minutes of auditory attention and 18 minutes of distractions. The EEG was collected with a 24-channel Smarting mobile EEG recording system.

The data are preprocessed according to the suggested methods in \cite{Biesmans2017}. The auditory stimulus is first split into subbands $y(t)$ with a gammatone filter bank. Each subband is then transformed according to the power law relation $|y(t)|^{0.6}$, which roughly corresponds to the relation between the intensity and the perceived loudness of sound. The envelopes of the transformed subbands are then extracted by applying a low-pass filter. All subband envelopes are finally added together with equal weights. The artefacts from the EEG are removed with a Wiener filter \cite{Somers2018}, where the same filter is applied to the entire EEG recording to avoid artificially introducing trial-specific signal statistics due to filtering. In the end, both the envelope and the EEG signal are filtered with a bandpass filter between 0.5 and 32 Hz and downsampled to 64 Hz \cite{Monesi2020}. 

The EEG and the stimulus are then cut into segments of various lengths (different decision window lengths will be tested). These segments are then processed following the steps Algorithm \ref{alg:UnsupCCA}.

\subsection{Experiments} \label{subsec:Experiments}
In a first experiment, we study the performance of the normal supervised CCA algorithm \eqref{eq:CCAOptimProblem}, the discriminative supervised CCA algorithm \eqref{eq:dCCAOptimProblem} and the unsupervised CCA algorithm proposed in Algorithm \ref{alg:UnsupCCA} using the default hyperparameters reported in Section \ref{subsec:hyperparam}. Since the dataset is heavily imbalanced, we will report the performance using the following metrics to paint a more accurate picture: accuracy, F1-score, the Receiver-operating characteristic (ROC) curve, and the area under the ROC curve (AUC). The attended class is considered to be the positive class when measuring the F1-score. 

Since the notion of a positive class is dubious in aAAD, and the accuracy can be misleading due to the class imbalance, we will focus on the AUC metric in all other experiments. Regardless of class imbalance, the AUC is always 0.5 for a random model and 1 for an optimal model. In general, the AUC increases when the projected 1-dimensional scores $y(n)$ are more separable.

In a second experiment, we will investigate the self-leveraging effect of the (non-discriminative) unsupervised CCA model in Algorithm \ref{alg:UnsupCCA} by training the CCA model with an initial set of labels that have a varying level of accuracy $p_i \in [0,1]$. This will test to what extent errors in the pseudo labels affect the performance of the algorithm. 

In a third experiment, we investigate the influence of the main novelties proposed in this paper on the final performance through an ablation study. This is done by gradually transforming the supervised CCA model into the unsupervised CCA model proposed in Algorithm \ref{alg:UnsupCCA}. This is done first by replacing the LDA classifier with the unsupervised MILDA classifier, then by replacing the supervised CCA model with an unsupervised CCA model trained with discrete labels ($1 if p(n)>0.5, 0 otherwise$), then by replacing the hard labels with soft labels, and finally by adding the maximally discriminative CCA model in the final iteration. 

In the fourth experiment, the influence of the window length $\tau$ of the segments $n$ is investigated. While larger window lengths typically lead to better performances \cite{Geirnaert2020, roebben_are_2024}, short window lengths are often interesting to achieve a better temporal resolution of when a subject starts and stops paying attention. 

Finally, the influence of the imbalance of the data set is investigated by randomly removing between 0\% and 90\% of the data from the attended class before training and testing the unsupervised algorithm. Since only data is removed from the attended class, the class imbalanced is gradually shifted in this experiment. However, at the same time the total amount of data is changed. To isolate these two effects, the results are compared to results where a proportional amount of data is removed from each class (i.e. where the class imbalance is kept constant). 

Each experiment is repeated for each subject separately. The supervised CCA models are always trained and tested using 10-fold cross-validation. The unsupervised models are run in a single batch, where the algorithm labels all data simultaneously. All significance tests between models are done with the paired Wilcoxon signed-rank test on the auc scores. 

\subsection{Model hyperparameters}
\label{subsec:hyperparam}
We use $K=2$ linearly independent decoders and encoders to construct the 2-dimensional feature vector $\bm{\rho}(n)$. Based on \cite{Cheveigne2020}, both the EEG signal and the envelope use lags up to 250ms. The EEG signal is also delayed by 200ms with respect to the speaker envelope. Since the sampling rate is 64Hz, this corresponds respectively to $L_x = L_s = 17$ lags for the EEG signal and speech envelope and a delay of $S=13$ samples between the speech envelope and EEG signal. Unless mentioned otherwise, all correlation features are computed using 10s-long non-overlapping windows. 

\section{Results and discussion}\label{sec:Results}
\subsection{Comparison of supervised and unsupervised CCA}
\begin{figure}
    \centering
    \includegraphics[width=\columnwidth]{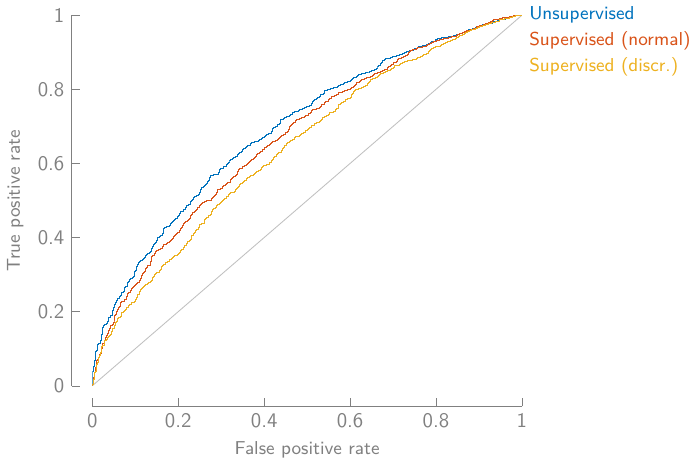}
    \caption{The average ROC-curves for the unsupervised model proposed in Algorithm \ref{alg:UnsupCCA} and the supervised models using normal and discriminative CCA. Remarkably, the unsupervised model outperforms the supervised model. The gray line represents chance level.}
    \label{fig:ROC-curves}
\end{figure}

The accuracy, F1-score and AUC for the unsupervised model proposed in Algorithm \ref{alg:UnsupCCA}, the supervised model using normal CCA \eqref{eq:CCAOptimProblem} and the supervised model using discriminative CCA \eqref{eq:dCCAOptimProblem} are shown in Table \ref{tab:performanceScores}. Their respective ROC-curves are shown in Figure \ref{fig:ROC-curves}. Remarkably, the unsupervised algorithm significantly outperforms both supervised models ($p=0.027$ and $p=0.0059$ for the normal and discriminative supervised models respectively). Furthermore, even though the discriminative CCA objective is more relevant for the aAAD classification problem than the normal CCA objective, the discriminative CCA model performs worse than the normal CCA model, though the difference is not significant ($p=0.084$). This is unexpected, especially because the inclusion of the discriminative objective in the unsupervised model does improve the performance, as will be shown in the ablation study in Section \ref{subsec:ablationStudy}.

A possible explanation is that the supervised model, which is cross-validated, does not generalise well on new data. This is not an issue for the unsupervised model, which iteratively re-trains itself on the same data it labels. This assumption is further reinforced in the ablation study, where the replacement of the LDA classifier with the unsupervised MILDA classifier significantly improves the performance of the supervised models. 

\begin{table}[]
    \centering
    \begin{tabular}{c|c c c}
         &                              Accuracy  &      F1-score &  AUC\\
         \hline
         Supervised (normal)            & $0.63 \pm 0.04 $ & $0.73 \pm 0.04$ & $0.67 \pm 0.06$  \\
         Supervised (discr.)            & $0.60 \pm 0.07 $ & $0.69 \pm 0.06$ & $0.64 \pm 0.07$  \\
         Unsupervised                   & $0.71 \pm 0.06 $ & $0.80 \pm 0.06$ & $0.70 \pm 0.06$
         \end{tabular}
    \caption{The accuracy, F1-score and AUC of the supervised CCA model based on \eqref{eq:CCAOptimProblem}, the discriminative supervised CCA model \eqref{eq:dCCAOptimProblem} and the unsupervised model proposed in Algorithm \ref{alg:UnsupCCA}. Remarkably, the unsupervised model outperforms the supervised models significantly.}
    \label{tab:performanceScores}
\end{table}

\subsection{Self-leveraging effect}
\begin{figure}
    \centering
    \includegraphics[width=\columnwidth]{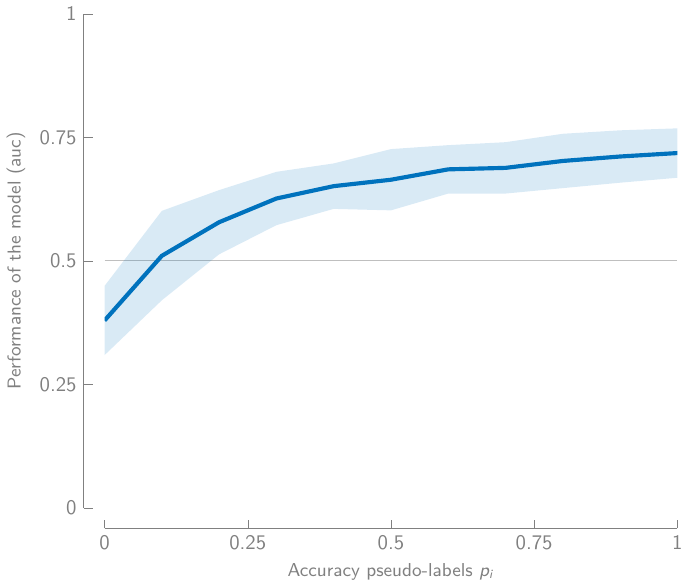}
    \caption{The average auc and standard deviation (shaded area) after training the model with pseudo-labels that are $p_i$ accurate. The model performs significantly better than chance at $p_i = 0.5$. This demonstrates that the model will naturally improve itself at discriminating between attended and unattended EEG. The gray line demonstrates the chance level.}
    \label{fig:updateCurve}
\end{figure}

In this experiment, we investigate the self-leveraging effect of the unsupervised algorithm by measuring its performance after training it with labels that are $p_i \in [0,1]$ accurate. As shown in Figure \ref{fig:updateCurve}, this self-leveraging effect is clearly present. When the model is initialised with random labels ($p_i=0.5$), it achieves an AUC of 0.66. This is much closer to the performance of a model trained with perfect labels (auc = 0.71) than chance level (auc = 0.5). Since the CCA model trained with random labels barely performs worse than the CCA model trained with perfect labels ($p_i = 1$), the algorithm converges very quickly to a point close to the theoretical optimum. In practice, the algorithm converges in only 2 to 3 iterations.  

Furthermore, even if the algorithm is trained to maximise the unattended correlations, rather than the attended correlations ($p_i = 0$), it performs close to chance level after 1 iteration, and will afterwards converge to a similar point after an additional 2 to 3 iterations. This means that the algorithm is capable to reach a good optimum from any intialisation, even in a worst-case scenario. 

\subsection{Ablation study}
\label{subsec:ablationStudy}
\begin{figure}
    \centering
    \includegraphics[width=\columnwidth]{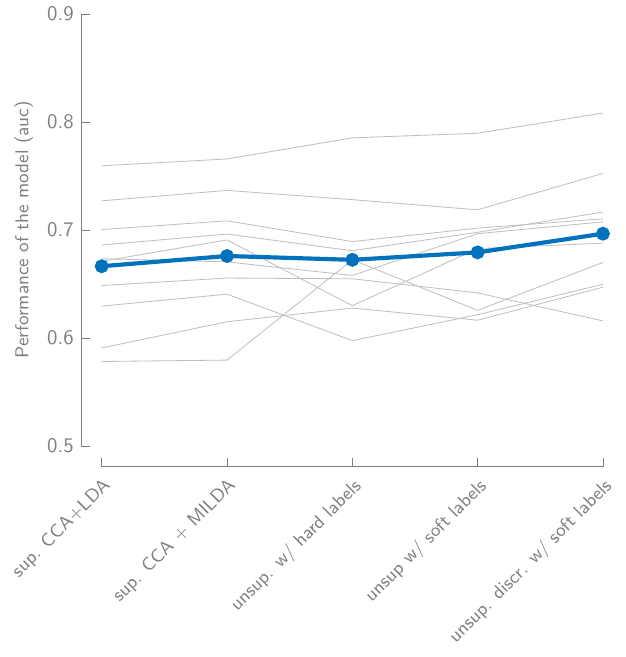}
    \caption{The ablation study of the proposed unsupervised model shows that the replacement of LDA by MILDA and the introduction of the discriminatory CCA model have the largest impacts on performance. Meanwhile, the transition from supervised to unsupervised CCA only leads to a minimal drop in performance. The blue line is the average AUC over all subjects, each gray line represents a subject.}
    \label{fig:ablation}
\end{figure}

Figure \ref{fig:ablation} shows how the auc changes as the model is gradually transformed from the supervised CCA + LDA model \eqref{eq:CCAOptimProblem} to the unsupervised algorithm proposed in Algorithm \ref{alg:UnsupCCA}. Replacing the LDA classifier with the unsupervised MILDA classifier significantly improves the auc with 1\% ($p=0.006$). Next, the supervised, cross-validated CCA model is replaced by an unsupervised CCA model that is iteratively re-trained using hard labels, similar to the updating scheme proposed in \cite{Geirnaert2021Unsup}. As expected, this leads to a minor, non-significant drop in auc ($-0.3\%, p=0.56$). This drop is immediately counter-acted if the hard labels are replaced by the soft labels $p(n)$, which estimate the probability that the subject is attentively listening to the story in segment $n$. This allows the model to put less importance on training samples where it is unclear whether the subject was paying attention or not. Finally, the introduction of the maximally discriminative CCA in the last iteration of the unsupervised algorithm further significantly improves the auc with $2\%$ ($p=0.027$). 

Although it is rather unexpected for unsupervised models to outperform supervised models, it does demonstrate their key strength. Unsupervised models are able to adapt to the numerous non-stationarities in EEG recordings by (re-)training the model on new data, whereas supervised models must be trained on a separate dataset and generalise on new data. 

Although this hypothesis does not explain all results. The supervised model improved when the LDA classifier was replaced by MILDA, while both supervised models (including the classifier) are cross-validated. A possible hypothesis is that MILDA performs better than LDA because it combines data with the prior (assumed) knowledge that the direction of maximal variance in the feature space is proportional to $\bm{\mu_\Delta}$, while LDA only relies on data. The reliance of MILDA on prior knowledge can be seen as a form of regularisation of the classifier, which is especially relevant when only a limited amount of feature vectors are available of a specific class (e.g. due to the inherent class imbalance in the dataset). 

\subsection{Decision window length}
Increasing the decision window length of test segments significantly improves the accuracy of the predictions at the cost of decision speed. Longer windows are thus not necessarily better, although the severity of this trade-off is application-dependent. We therefore analyse the performance of the unsupervised model for various decision window lengths $\tau$, ranging from 1s to 30s. 

\begin{figure}[ht]
    \centering
    \includegraphics[width=1.1\columnwidth]{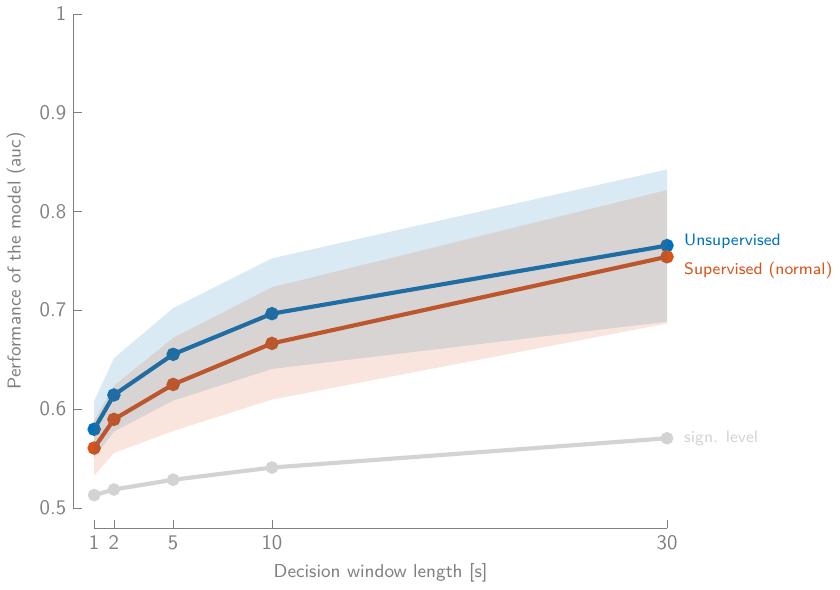}
    \caption{As expected, the unsupervised model performs better as the correlation features are estimated on longer window lengths. Nevertheless, it also performs better than chance on small window lengths. The solid line represents the average and the shaded area represents the standard deviation. The gray line is the significance level.}
    \label{fig:ptaucurve}
\end{figure}

This trade-off between temporal resolution and performance is clearly visible in Figure \ref{fig:ptaucurve}. The algorithm performs better as longer correlation windows are used. This is in line with previous findings \cite{OSullivan2014, Geirnaert2020, DeCheveigne2018, roebben_are_2024}. The unsupervised model significantly outperforms the supervised model on all window lengths but the largest window length ($p=0.02$ if $\tau \leq 10$s, $p=0.38$ if $\tau=30$s). Both models always perform significantly better than chance. 

\subsection{Class imbalance}
The dataset used in this dataset has an inherent class imbalance where about 75\% of all data belongs to the attention class. To verify whether the proposed algorithm also performs well when the classes are balanced, or imbalanced in the other direction, we gradually remove up to 90\% of the data from the attended class while keeping the unattended class unchanged. As a consequence, the fraction of data belonging to the attended class gradually shifts from 75\% to 23\%. This result is compared to a control experiment where a proportional amount of data is removed from each class, such that the effect of removing data can be separated from the effect of the class imbalance. 

\begin{figure}[t]
    \centering
    \includegraphics[width=\columnwidth]{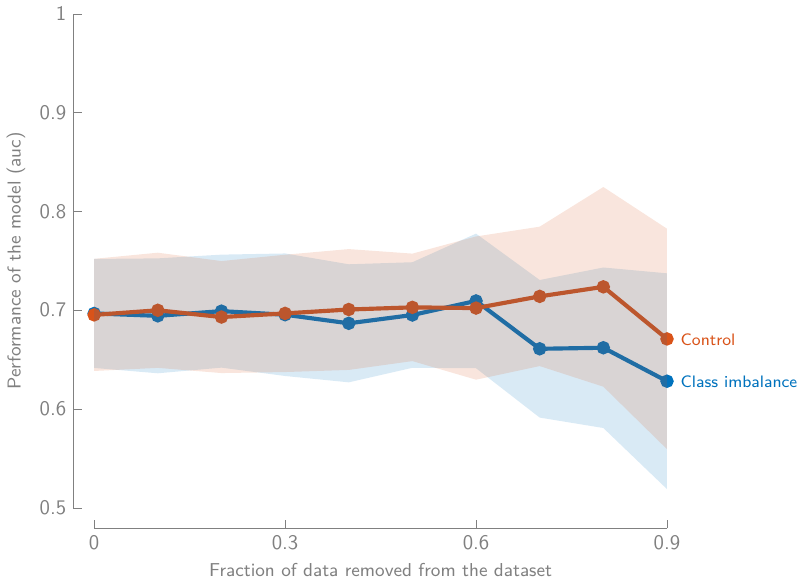}
    \caption{The performance of the unsupervised model in function of the fraction of the dataset that is removed. For class imbalance, the data is uniquely removed from the attended class, such that the class imbalance is changed. For control, the data is proportionally removed from both classes, such that the class imbalance stays the same. Solid lines represent the averages and the shaded area the standard deviation.}
    \label{fig:classImbalance}
\end{figure}

As shown in Figure \ref{fig:classImbalance}, the class imbalance has initially little influence on the performance of the model. Once more than 60\% of the attended class is removed, the performance gradually drops, with a difference in auc of about $7\% \pm 9\%$  compared to the full dataset. This is mostly caused by the unsupervised CCA algorithm that is iteratively computed. This model solely relies on data from the attended class, which is extremely limited when 90\% of that data is removed. Furthermore, there are much more unattended than attended segments in that case. These unattended segments will, therefore, have a significantly larger influence on the estimation of $R_{xs}$, which diminishes the self-leveraging effect on which the CCA algorithm relies. 

On the other hand, it is quite surprising that the control experiment, where data was proportionally removed from both classes, does not show a significant influence of the amount of data that is removed on the performance of the algorithm. At the limit, the unsupervised algorithm is limited to 7 minutes of data, yet does not perform significantly worse than when the full dataset is used. This makes the unsupervised algorithm especially suited for adaptive use cases, where the models are constantly retrained on new, incoming data.

\subsection{Implications for online wearable devices}
Even though we evaluated the unsupervised algorithm in this paper offline with a batch of earlier recorded data, it can be applied in an online, wearable setting. In such a setting, the newly recorded speech and EEG segments are added to the existing dataset to update the model. At first sight, such an algorithm seems to require an unrealistic amount of computing power and memory for online wearable devices. However, on closer inspection, the actual memory and computational requirements are barely larger than those of a supervised, static algorithm. 

The entire model relies on the estimation of the CCA parameters $R_{xx}$, $R_{ss}$, $R_{xs}$, the LDA parameters $\Bar{\bm{\mu}}$, $\Bar{\Sigma}$ and the GMM parameters $\mu_+$, $\sigma_+$ and $\sigma_-$. Since all these parameters are estimated by computing some average over all segments, they can be updated recursively using the formula $X'=\alpha X + X(n)$, with $X$ the old parameter, $X'$ the updated parameter, $X(n)$ the segment specific parameter and $\alpha \in [0,1]$ a forgetting factor. It thus suffices to know the segment specific parameters, i.e. to know the correlation matrices $R_{xx}(n)$, $R_{ss}(n)$, $R_{xs}(n)$, the feature vector $\bm{\rho}(n)$, its projection $y(n) =\Vec{w}^T\bm{\rho}(n)$ and the variances $(\bm{\rho}(n)-\Bar{\bm{\mu}})(\bm{\rho}(n)-\Bar{\bm{\mu}})^T$ and $(y(n)-\mu_{+/-})^2$ of the new segments.

Apart from the variances $(\bm{\rho}(n)-\Bar{\bm{\mu}})(\bm{\rho}(n)-\Bar{\bm{\mu}})^T$ and $(y(n)-\mu_{+/-})^2$, which are cheap to compute, all these parameters are already estimated when the new pair of EEG and speech segments is classified, as shown in Algorithm \ref{alg:UnsupCCA}. Updating the model parameters thus barely requires any additional computation compared to classifying new segments. Updating the actual filters $\Vec{d}$, $\Vec{e}$ and $\Vec{w}$ still requires some additional computation. However, this costs significantly less than, for instance, estimating the correlation matrices as long as the filters are not updated too frequently. 

The recursive implementation of the unsupervised model discussed above thus allows us to (re-)train the unsupervised model on a large amount of training data while a person is using the wearable device. However, increasing the amount of training data does come with an important drawback. As a model is trained on more data, it takes more time to adapt to sudden changes such as the failure of an electrode or a vastly different listening environment. This trade-off should be studied in future work where speech decoders are used in changing environments.

\section{Future work and conclusion}
\label{sec:Conclusion}
We have successfully developed an adaptive, unsupervised model that decodes from an EEG signal whether a subject is actively listening to an auditory stimulus. The model uses a CCA-based feature extraction module, followed by an unsupervised classifier that is equivalent to LDA and a GMM for thresholding. To this end, we have demonstrated a self-leveraging effect of CCA, and we have incorporated minimally informed LDA classifier (called MILDA) that is independent of the class labels. The model is demonstrated to converge after two to three iterations. After convergence, it performs significantly better than its supervised counterpart without requiring any supervised training or initialisation. The cause of this remarkable result was closely investigated through an ablation analysis. 

We have demonstrated that the algorithm successfully works at various levels of class imbalance, and even when the amount of available data is extremely limited. Since the algorithm is also relatively cheap to update on new data, it is well suited for use in adaptive contexts where the models continuously adapts to the changing statistics of the incoming data. 

\section*{Acknowledgement}
We would like to thank Elly Brouckmans and Linsey Dewit-Vanhaelen for collecting the dataset that was used in this paper.

\bibliographystyle{IEEEtran}
\bibliography{IEEEabrv,references}
\end{document}